	\def\matsymbol#1{{\bf {#1}}}
	\newcommand{\be}{\begin{equation}}
        \newcommand{\ee}{\end{equation}}
        \newcommand{\hf}{{1\over 2}}
        \newcommand{\ba}{\begin{eqnarray}}
        \newcommand{\eps}{\epsilon}
        
        \newcommand{\tensor}{\otimes}
        \newcommand{\maps}{\colon}
        \newcommand{\ea}{\end{eqnarray}}
	\newcommand{\lb}{\lbrack\!\lbrack}
	\newcommand{\rb}{\rbrack\!\rbrack}
	\newcommand{\fps}{\lb \eps \rb}

	\renewcommand{\L}{{\cal L}}

        \newcommand{\C}{{\matsymbol C}}
	\newcommand{\CS}{{\matsymbol C}S}
	\newcommand{\CB}{{\matsymbol C}B}
	\newcommand{\CGB}{{\matsymbol C}GB}

	\renewcommand{\H}{{\bf H}}
        \newcommand{\qed}{\hskip 3em \hbox{\BOX} \vskip 2ex}
        \newcommand{\BOX}{\hbox {$\sqcap$ \kern -1em $\sqcup$}}
 	\newcommand{\et}{\hspace{-0.08in}{\bf .}\hspace{0.1in}}
	\newcommand{\tr}{{\rm tr}}

	\newtheorem{lemma}{Lemma}
	\newtheorem{corollary}{Corollary}
	\newtheorem{theorem}{Theorem}
	
	\documentstyle[12pt]{article}
\begin{document}

	\begin{center}
	{\bf Link Invariants of Finite Type and \\}
	{\bf  Perturbation Theory \\}
	\vspace{0.5cm}
	{\em John C. Baez\\}
	\vspace{0.3cm}
	{\small Department of Mathematics \\
	Wellesley College\\
	Wellesley, Massachusetts 02181\\
	(on leave from the University of California at Riverside)\\ }
	\vspace{0.3cm}
	{\small  July 13, 1992}
	\vspace{0.5cm}
	\end{center}

\begin{abstract}
The Vassiliev-Gusarov link invariants of finite type are known to be
closely related to perturbation theory for Chern-Simons theory.  In
order to clarify the perturbative nature of such link invariants, we
introduce an algebra $V_\infty$ containing
elements $g_i$ satisfying the usual braid group relations and
elements $a_i$ satisfying $g_i - g_i^{-1} = \eps a_i$, where $\eps$ is
a formal variable that may be regarded as measuring the failure of
$g_i^2$ to equal 1.   Topologically, the elements $a_i$ signify
crossings.
We show that a large class of link invariants of finite type are in
one-to-one correspondence with homogeneous Markov traces on
$V_\infty$.
We sketch a possible application of link invariants of finite type to
a manifestly diffeomorphism-invariant perturbation theory for quantum
gravity in the loop representation.
\end{abstract}

\section{Introduction}

The manner in which the braid group $B_n$ takes the place of the
symmetric group $S_n$ in the representation theory of quantum groups
is by now well known.
Recall that the braid group $B_n$ has a presentation with generators
$s_i$, $1 \le i < n$, and relations
\ba    s_i s_j &=& s_j s_i \qquad\qquad\qquad |i-j| > 1, \nonumber\cr
s_is_{i+1}s_i &=& s_{i+1}s_i s_{i+1} .\nonumber\ea
The symmetric group $S_n$ is the quotient of $B_n$ by the further
relations $s_i^2 = 1$.
If $G$ is a semisimple Lie group, then the corresponding quantized
enveloping algebra
$U_q{\bf g}$   is a deformation of the universal enveloping
algebra $U{\bf g}$  as Hopf algebras.     Given any
representation of the group $G$,
there is a natural action of the group algebra
$\CS_n$ as intertwining operators on the representation $E^{\tensor n}$.
Similarly, given any representation $E$ of $U_q{\bf g}$ there
is a natural action of $\CB_n$
as intertwining operators on $E^{\tensor n}$.

Naively, one might be led to hope that $\CB_n$ is
a kind of deformation of $\CS_n$.  While this is not true according to
the standard definition - after all,
$\CB_n$ is infinite-dimensional while $\CS_n$ is finite-dimensional -
we show here that some sense can be made of this idea.
Roughly speaking, one can form an algebra $V_n$ over $\C[\eps]$, where
$\eps$ is a formal variable, by adjoining to $\CB_n$ elements
$a_i$, $1 \le i < n$, such that
\[                  s_i - s_i^{-1} = \eps a_i  .\]
The parameter $\eps$ should be thought of as measuring the
failure for $s_i^2$ to equal $1$.  Setting
$\eps$ equal to any nonzero constant gives the algebra $\CB_n$, while
setting $\eps = 0$ gives an algebra containing $\CS_n$.
More generally, for any $d \ge 0$,
the quotient algebra $V_n/\langle \eps^{d+1} \rangle$
should be regarded as a $d$th-order perturbative approximation to $\CB_n$.

These quotients are closely related to the theory of link invariants of
finite type, as developed by Vassiliev, Gusarov, Birman, Lin,
Bar-Natan, and others \cite{Bar-Natan,BL,Gus,Lin,Vass}.
The basic idea here is that one can canonically extend an invariant
$\L$ of
oriented links to an invariant of generalized links
admitting nice self-intersections (transverse double points)
by means of the rule
\[      \L(L_+) - \L(L_-) = \eps \L(L_\times) ,\]
Here $L_+$ is as in Figure 1, $L_-$ is as in Figure 2, and $L_\times$
is as in Figure 3, with the strands oriented so as to be pointing
{\it downwards}.
\begin{center}
\setlength{\unitlength}{0.0125in}%
\begin{picture}(45,60)(10,770)
\thicklines
\put( 10,830){\line( 3,-4){ 17.640}}
\put( 36,794){\line( 4,-5){ 19.122}}
\put( 55,830){\line(-3,-4){ 45}}
\end{picture}
\vskip 1em
Figure 1.
\end{center}
\begin{center}
\setlength{\unitlength}{0.0125in}%
\begin{picture}(45,60)(10,770)
\thicklines
\put( 55,830){\line(-3,-4){ 18}}
\put( 28,794){\line(-3,-4){ 18}}
\put( 10,830){\line( 3,-4){ 45}}
\end{picture}
\vskip 1 em
Figure 2.
\end{center}
\begin{center}
\setlength{\unitlength}{0.0125in}%
\begin{picture}(45,60)(10,770)
\thicklines
\put( 55,830){\line(-3,-4){ 45}}
\put( 10,830){\line( 3,-4){ 45}}
\end{picture}
\vskip 1em
Figure 3.
\end{center}
Those invariants vanishing on
all generalized links with more than $d$ self-intersections are said
to be of degree $d$.   The space of link invariants
of degree $d$ can be regarded as a
$d$th-order approximation to the dual of the space with basis given by
isotopy classes of links.
If one takes one of the $\C(q)$-valued link invariants
derived from quantum group representations by the procedure of
Turaev \cite{Tur}, sets $q = \exp(\eps)$ to obtain a formal power
series in $\eps$, and takes the coefficient
of $\eps^d$, one obtains a link invariant of degree $d$.
More generally, we show that there is a one-to-one correspondence
between a large class of link invariants of degree $d$
and Markov traces $\tau \maps V_\infty \to \C[\eps]$ that are
``homogeneous of degree $d$'' in a certain sense.  Such Markov traces
may be thought of as defined on $V_\infty/\langle \eps^{d+1}\rangle$.

The connection between link invariants of finite type
and {\it physical} perturbation theory is presently clearest
in the context of Chern-Simons theory with
semisimple gauge group $G$.  Here the action is given by
\[   S = {k\over 4\pi} \int_{S^3} \tr (A \wedge dA + {2\over 3}A \wedge A
\wedge A) ,\]
where $A$ is a $G$-connection and the level $k \ge 0$ is an integer.
{}From the work of Witten \cite{Witten}
and others it is clear that, for example, the Jones polynomial $V_L(q)$
may be obtained from the vacuum
expectation values of Wilson loops in Chern-Simons theory with $G =
SU(2)$ and $q = \exp(2\pi i/(k+2))$.
The coefficient of the $\eps^d$ term of the
Jones polynomial, an invariant of finite degree, should thus be
calculable by a $d$th-order perturbation expansion in Chern-Simons
theory.   This has been pursued by Cotta-Ramusino {\it et al}, Smolin,
and others \cite{Guad,Smolin1}.
In particular, the relation to knot invariants of finite type
has been studied by Bar-Natan \cite{Bar-Natan}, who
dealt with an arbitrary classical Lie group,
and by abstraction obtained
a general combinatorial scheme for constructing knot invariants of
finite type.

Link invariants of finite
type may also be expected to play a role in a novel
perturbation theory for $4$-dimensional quantum
gravity.   In the loop representation of quantum gravity developed by
Rovelli  and Smolin \cite{RS}, states are described by linear
combinations of isotopy classes of framed unoriented links (or tangles),
possibly admitting self-intersections.
We briefly comment on the relation
between Chern-Simons perturbation theory, link invariants of finite
type, and perturbative quantum gravity in the final section of this paper.

The author thanks Dror Bar-Natan, J\'ozef Przytycki, and
Stephen Sawin for useful discussions concerning link invariants of
finite type, and thanks Micheal Weiss for help with drawing the figures.
\eject

\section{The Vassiliev Algebra}

Let the {\it generalized braid monoid}, $GB_n$, denote
the monoid with generators
$g_i, g_i^{-1}, a_i$, $1 \le i < n$, and relations
\ba    [g_i, g_j] &=& \;[a_i, a_j]\; = \;[a_i, g_j]\; = 0
\qquad\qquad\qquad |i-j| > 1, \nonumber\cr
g_ig_i^{-1} &=& g_i^{-1}g_i \;=\; 1  , \nonumber\cr
a_i g_i &=& g_i a_i  ,\nonumber\cr
g_ig_{i+1}g_i &=& g_{i+1}g_i g_{i+1}, \nonumber\cr
g_{i+1}a_ig_{i+1}^{-1} &=&  g_i^{-1}a_{i+1}g_i  ,\nonumber\cr
g_{i+1}^{-1}a_ig_{i+1} &=&  g_i a_{i+1}g_i^{-1} .\nonumber\ea
In pictures of generalized braids, $g_i$ represents a right-handed
crossing and $g_i^{-1}$ represents a left-handed crossing of the $i$th and
$(i+1)$st strands, as in Figures 1 and 2.  The element $a_i$ represents an
intersection of the $i$th and $(i+1)$ strands, as in Figure 3. The
relations above express topological facts about generalized braids,
which admit intersections as well as crossings; the reader is strongly
encouraged to draw these relations.
The generalized braid monoid appears in the work of Kaufmann
\cite{Kauffman}, as well as in the work of
Br\"ugmann, Gambini and Pullin \cite{BGP,G} on quantum gravity.

Let $\CGB_n$ denote the monoid
algebra of the generalized braid monoid.
We define the {\it Vassiliev algebra}, $V_n$, to be the
quotient of $\CGB_n \tensor \C[\eps]$ by the ideal generated by the
elements
\[g_i - g_i^{-1} =  \eps a_i  .\]
It is clear that there is a homomorphism
\[     v \maps \CB_n \to V_n \]
given by $s_i \mapsto g_i$.
The basic properties of this homomorphism are as
follows.

\begin{lemma}\label{lem1}\et  Let $\C(\eps)$ denote the algebra of Laurent
polynomials in $\eps$.  Then
\[v \tensor 1 \maps \CB_n \tensor
\C(\eps) \to V_n \tensor_{\C[\eps]} \C(\eps)\]
is an isomorphism.
\end{lemma}

Proof - The inverse is
given by $g_i \mapsto s_i$, $a_i \mapsto \eps^{-1}(s_i - s_i^{-1})$.
\qed

\begin{corollary} \label{cor1}\et   Let $j \maps V_n \to V_n/ \langle
\eps - x\rangle$ denote the quotient map.  The map
\[        j\circ v \maps \CB_n \to V_n /\langle \eps - x\rangle \]
is an isomorphism if $x \in C$ is nonzero, while if $x = 0$ it factors
through $\CS_n$.
\end{corollary}

Proof - The composite $j\circ v$ is an isomorphism for $x \ne 0$ by
Lemma \ref{lem1}, while if $x = 0$, $s_i^2 = 1$ in $V_n/\langle
\eps\rangle$, so $j\circ v$ factors through $\CS_n$.  \qed

\begin{corollary} \label{cor2}\et  The homomorphism $v \maps \CB_n \to
V_n$ is one-to-one.  \end{corollary}

Proof - This is immediate from Lemma \ref{lem1}.  \qed

We conclude this section with a word on the universal role the
Vassiliev algebra plays in the context of braided tensor categories.
It is clear that, given any object $E$ in a $\C[\eps]$-linear strict
braided monoidal category, if $R \equiv R^{-1} \bmod
\eps$, where $R \maps E \tensor E \to E \tensor E$ is the
braiding, then there is a canonical action of $V_n$ as endomorphisms of
$E^{\tensor n}$.  In particular, this applies to the category of
quantum group representations, where we write $q = \exp(\eps)$.

\section{Link Invariants of Finite Type}

By a link invariant we will always mean an ambient isotopy invariant
of oriented links.
It is easy to see that any $\C$-valued link invariant $\L$ uniquely
extends to a $\C(\eps)$-valued invariant of generalized
links admitting transverse double points, which we also call
$\L$, by means of the rule
\[       \L(L_+) - \L(L_-) = \eps\L(L_\times) ,\]
where $L_+$, $L_-$, and
$L_\times$ denote link diagrams with a right-handed crossing, a
left-handed crossing, and an intersection, respectively, at a given
point, the rest of the diagrams being the same.   We
define a $\C$-valued
link invariant to be of {\it degree $d$} if it vanishes on all generalized
links with $d+1$ or more self-intersections.
A link invariant of degree $d$ for some $d$ is said to be of {\it
finite type}.

For all $n$ there are algebra inclusions $V_n \hookrightarrow V_{n+1}$
and $\CB_n \hookrightarrow \CB_{n+1}$.
Let $V_\infty$ and $\CB_\infty$ denote the inductive limits of the
algebras $V_n$ and $\CB_n$, respectively, and let
\[    v \maps \CB_\infty \to V_\infty \]
denote the inductive limit of the maps $v \maps \CB_n \to V_n$.
We define a {\it Markov
trace} on $V_\infty$ to be a $\C[\eps]$-linear map $\tau \maps V_\infty
\to E$, where $E$ is some $\C[\eps]$-module, satisfying
\[           \tau(xy) = \tau(yx)  \]
for all $x,y \in V_\infty$, and for some fixed $z \in \C$
\[           \tau(g_n^{\pm 1} x) = z\tau(x)  \]
for all $x \in V_n \subset V_\infty$.
A similar definition is standard for Markov traces $\tr \maps
\CB_\infty \to \C$.  We say that a Markov trace $\tau \maps
V_\infty \to \C[\eps]$ is {\it homogeneous of degree $d$} if for every
$x \in \CB_\infty$, $\tau(v(x))$ is homogeneous of degree $d$ as a
polynomial in $\eps$.

For any braid $x \in B_n$, let $\hat x$ denote its closure.  Of
course, given $x \in B_\infty$, $\hat x$ depends on a choice of
$n$ such that $x \in B_n$.   Given a link $L$, let $L \cup \circ$
denote the distant union of $L$ with the unknot.

\begin{theorem}\et There is a one-to-one correspondence between
$\C$-valued link invariants $\L$ of degree $d$
such that for some $z \ne 0$ and all links $L$,
\be        \L(L \cup \circ) = z^{-1} \L(L),       \label{circ}  \ee
and Markov traces $\tau \maps V_\infty \to \C[\eps]$
that are homogeneous of degree $d$.
The invariant $\L$ determines the trace $\tau$, and conversely, by the
property that
\be     \tau(v(x)) = \eps^d z^{n-1}\L(\hat x)      \label{CORR} \ee
for $x \in B_n$.
\end{theorem}

Proof -
Let $E$ be a vector space, and suppose $\L$ is a $E$-valued link
invariant satisfying equation (\ref{circ}).   Then by
Markov's theorem \cite{Birman} there is a
Markov trace $\tr \maps \CB_\infty \to E$, given by
\be               \tr(x) = z^{n-1} \L(\hat x)    \label{corr}  \ee
for all $x \in B_n$.  Note in particular that $\tr$ is well-defined
because if $y \in \CB_{n+1}$ is the image of $x \in \CB_n$ under the
inclusion $\CB_n \hookrightarrow \CB_{n+1}$, then
\[  \tr(y) = z^n\L(\hat y) = z^n\L(\hat x \cup \circ) =
z^{n-1}\L(\hat x) = \tr(x) .\]
Moreover $\tr$ is a trace because $\widehat{xy} = \widehat{yx}$ for
all $x,y \in B_n$, and $\tr$ has the Markov property:
\[   \tr(s_n^{\pm 1}x) = z^n \L((s_n^{\pm 1}x)\widehat{\;})
= z^n \L(\hat x) = z\, \tr(x)  .\]
In fact, Markov's theorem implies that equation (\ref{corr})
gives a one-to-one correspondence
between $E$-valued link invariants satisfying equation (\ref{circ})
and Markov traces $\tr \maps \CB_\infty \to E$.

Now let $\L$ be a $\C$-valued link invariant of degree $d$ satisfying
equation (\ref{circ}).   Then there is a unique Markov trace
$\tr \maps \CB_\infty \to \C$ given by equation (\ref{corr}).
We claim that there exists a Markov trace
$\tau_0 \maps V_\infty \to \C(\eps)$ such that
\[        \eps^d \tr = \tau_0   v .\]
To see this, note that by Lemma \ref{lem1} there is a Markov trace
$\tilde\tau_0 \maps V_\infty \tensor_{\C[\eps]} \C(\eps)
\to \C(\eps)$ such that
\[          \eps^d (\tr \tensor 1) = \tilde\tau_0 (v \tensor 1)\]
as maps from $\CB_\infty \tensor \C(\eps)$ to $\C(\eps)$.
Let $\tau_0$ denote the restriction of $\tilde \tau_0$ to
$V_\infty$, which may be regarded as a subalgebra of $V_\infty
\tensor_{\C[\eps]} \C(\eps)$.  It follows that $\eps^d \tr =
\tau_0 v$, as desired.

Note that $\tau_0(x) = c\eps^{d-\ell}$ for some $c \in \C$ if $x \in
V_\infty$ is a product of $\ell$ elements of the form $a_i$ and
arbitrarily many of the form $g_i$.   Moreover, if $\ell > d$
then $\tau_0(x)$ vanishes, since $\L$ is of degree $d$.  It follows that
\[        \tau_0(x) = \sum_{i=0}^d c_i \eps^i,  \]
with $c_i \in\C$, if $x$ is a product of elements of the form $a_i$
and $g_i$.   It follows that $\tau_0$ factors through a map
$\tau \maps V_\infty \to C[\eps]$.
It is easy to check that $\tau_0$, hence $\tau$, is a Markov trace
that is homogeneous of degree $k$.  Moreover,
equation (\ref{CORR}) holds by construction.

Conversely, suppose that $\tau \maps V_\infty \to C[\eps]$
is a Markov trace homogeneous of degree $d$.  We may define a Markov trace
$\tr \maps \CB_\infty \to \C$ by
\[        \tau v = \eps^d \tr .\]
Associated to this trace there is a link invariant $\L$ satisfying equation
$\ref{circ}$, given by equation (\ref{corr}).  We claim that $\L$ is
an invariant of degree $d$.
For this, we need an analog of Alexander's theorem for
generalized links.   Given an element $x \in GB_n$, we may form a
generalized link $\hat x$, the {\it closure} of $x$, in a manner
analogous to the usual closure of a braid.

\begin{lemma}\et  For every generalized link $L$, for some $n$
there is an element $x \in GB_n$ such that $\hat x$ is ambient
isotopic to $L$.  \end{lemma}

Proof - We omit the proof, as it is similar to the usual
proof of Alexander's theorem \cite{Birman}, but would be quite
long with all the details included.   \qed

Now let $L$ be a generalized link with $\ell$ self-intersections, and
let $x_0 \in GB_n$ be such that $\hat x_0$ is ambient isotopic to $L$.
Let $x_1$ be the image of $x_0$ in $V_\infty$.
Then
\[     x_1 = y_1a_{i_1}y_2a_{i_2} \cdots a_{i_\ell}y_{\ell + 1}  \]
where the elements $y_i$ are products of elements $g_j, g_j^{-1}$.
Define $x \in \CB_\infty$ by
\[     x = y_1(g_{i_1} - g_{i_1}^{-1})y_2
 \cdots (g_{i_\ell} - g_{i_\ell}^{-1})y_{\ell + 1} . \]
Note that
\[         \eps^d \tr(x) = \tau(v(x))  = \eps^\ell
\tau(x_1) . \]
Since $\tr(x) \in \C$ and $\tau(x_1) \in \C[\eps]$, so
if $\ell > d$ we must have $\tr(x) = 0$.
By construction,
\[      \L(L) = z^{1-n}\tr(x) ,\]
so it follows that $\L(L) = 0$.

Now let us show that the invariant $\L$ uniquely determines the trace
$\tau$, and vice versa, by equation (\ref{CORR}).   Since every link
is the closure of some braid, $\L$ is determined by $\tau$.
Conversely, $\L$ determines $\tau$ on the image of $v$, since $v$ is
one-to-one by Corollary \ref{cor2}.   It then follows by the
$\C[\eps]$-linearity of $\tau$ that $\tau$ is determined on all of
$V_\infty$, since $\eps a_i = s_i - s_i^{-1}$.
 \qed

The reader may be puzzled by the fact that every link invariant of
degree $d$ is obviously of degree $d+1$, while a normalized Markov
trace $\tau \maps V_\infty \to \C[\eps]$ that is homogeneous of degree
$d$ is definitely {\it not} homogeneous of degree $d+1$.
The point is that $\eps \tau$ will be a normalized Markov
trace homogeneous of degree $d+1$.

Note that to reconstruct the link invariant coming from a normalized
Markov trace  $\tau \maps V_\infty \to \C[\eps]$ that is homogeneous
of degree $d$, it suffices to know the composite of $\tau$ with the
quotient map $\C[\eps] \mapsto \C[\eps]/\langle \eps^{d+1} \rangle$,
which may be regarded as a Markov trace
\[      \tau \maps V_\infty/\langle \eps^{d+1}\rangle \to
\C[\eps]/\langle \eps^{d+1} \rangle .\]

As an illustration of the theorem,
let $\tr_0 \maps \CB_\infty \to
\C(q)$ be one of the traces obtained from quantum group
representations by Turaev's procedure.  Writing $q = \exp(\eps)$, we
regard $\tr_0$ as having values in $\C\fps$.
For some invertible $z \in \C\fps$,
\[     \tr_0(s_n^{\pm 1} x) = z\,\tr_0(x)  \]
for all $x \in \CB_n$.  Since $z \notin \C$, the theorem above is not
directly applicable.  However, we may define a new Markov trace $\tr \maps
\CB_\infty \to \fps$ with
\[      \tr(s_n^{\pm 1} x) = \tr(x)  \]
for all $x \in \CB_n$ by setting
\[      \tr(x) = z^{m-n} \tr_0(x)  ,\]
for $x \in B_n$, where $m$ denotes the number of components of $\hat x$.
There is a unique Markov trace $\tau \maps V_\infty \to \C\fps$ such that
\[    \tau v = \tr .\]
Moreover, we may write
\[       \tau = \sum_{d= 0}^\infty \tau_d  \]
where $\tau_d \maps V_\infty \to \C[\eps]$ is a Markov trace
homogeneous of degree $d$, with $\tau_d(g_n^{\pm 1}x) = \tau_d(x)$
for all $x \in V_n$.  By the theorem, each such trace gives a
link invariant $\L_d$ of degree $d$, such that
\[            \L_d(\hat x) = \eps^{-d} \tau_d(v(x)) \]
for $x \in B_n$.

It is also important that the space of Markov traces $\tau \maps
V_\infty \to \C[\eps]$ that are homogeneous of degree $d$ is
finite-dimensional.  This may be seen by graph-theoretic reasoning
along the lines of Birman and Lin \cite{BL}, Bar-Natan
\cite{Bar-Natan}, and Stanford \cite{Stanford}.
Suppose, for example, that we fix the value of $z$.
Then if $d = 0$,
$\tau$ is determined by its value on $1$.   If $d = 1$,
$\tau$ is determined by its value on $1, a_1,$ and $a_1s_1$.  If
$d = 2$, $\tau$ is determined by its value on $1, a_1, a_1s_1,
a_1^2, a_1^2s_1, a_1a_2, a_1a_2s_1s_2, a_1a_3, a_1a_3s_1,$
and $a_1a_3s_1s_3$.

\section{Connections to Physics}

The Vassiliev algebra formalism makes clear that link invariants of
finite type arise from a sort of ``topological perturbation
theory.''  We have pointedly used the symbol $\eps$ in our paper,
instead of the suggestive letter $h$ (common
in the quantum group literature), because the physical interpretation
of this sort of perturbation theory is an
interesting issue.

In $SU(n)$ Chern-Simons theory one makes the identification
\[     \eps = {2\pi i\over k + n  } . \]
The limit $\eps \to 0$ thus corresponds to $k
\to \infty$, which may regarded either as a weak coupling limit or
as a classical limit.    In the weak coupling approach \cite{Witten} we
may write $A = A_0 + k^{1/2} B$, where
$A_0$ is a flat connection, and obtain
\[   S = kI + {1\over 8\pi} \int \eps^{abc}
 \tr (B_a D_b B_c + {2\over 3}k^{-\hf} B_a[B_b,B_c]),\]
where $I$ is the Chern-Simons invariant of $A_0$ and $D$ denotes the
covariant derivative with respect to $A_0$.    Here we see that the
$k \to \infty$ limit is closely related to a deformation of the Lie
algebra ${\bf su}(n)$ to an abelian Lie algebra by scaling the bracket.
Alternatively, since $k$ appears
where one would expect a factor of $\hbar^{-1}$ in the partition function
\[   Z = \int {\cal DA} \exp\left( {ik\over 4\pi} \int \tr (A \wedge
dA + {2\over 3}A \wedge A \wedge A\right), \]
one may also regard $k \to \infty$ as a classical limit.  This is
consistent with the weak coupling interpretation, of course, since the
classical solutions of Chern-Simons theory are flat connections.

In quantum gravity the two obvious limits to consider are the
classical limit $\hbar \to 0$ and the $G \to 0$ limit, where $G$ is
Newton's gravitational constant.  The latter has been considered in the
loop representation of Euclidean quantum gravity by Smolin
\cite{Smolin2}, with the aim of developing a new perturbation theory
for quantum gravity which is manifestly diffeomorphism-invariant at
every order, as opposed to perturbation about a flat background spacetime.
In this approach an $SU(2)$ connection is
a key dynamical variable \cite{Ash}, and Smolin shows that the role of
$G$ is to scale
the Lie bracket in ${\bf su}(2)$.   The analogy with the $k \to \infty$
limit of Chern-Simons theory is no accident, since
states of quantum gravity may be obtained
from Wilson loops in $SU(2)$ Chern-Simons theory \cite{Kodama}.
In this connection, the
extension of the Jones polynomial to generalized links by Br\"ugmann,
Gambini and Pullin \cite{BGP} is quite intriguing.

It is thus natural to
hope that link invariants of finite type will play an important role
in the $G \to 0$ limit of quantum gravity, or similar perturbation
expansions in more general tangle field theories \cite{Baez}.
Roughly, we may expect that the true physical Hilbert space $\H$
is an inverse limit of spaces $\H_d$, where two states in $\H$ are
identified in $\H_d$ if they cannot be distinguished by link (or tangle)
invariants of degree $d$.

 \end{document}